\documentstyle[prl,aps,preprint]{revtex}
\begin{document}
\draft
\title{Late Time Tail of Wave Propagation on Curved Spacetime}

\author{E.S.C. Ching${}^{(1)}$, P.T. Leung${}^{(1)}$,
W.M. Suen${}^{(2)}$, and K. Young${}^{(1)}$}

\address{${}^{(1)}$Department of Physics,
The Chinese University of Hong Kong, Hong Kong}
\address{${}^{(2)}$Department of Physics, Washington University,
St Louis, MO 63130, U S A}

\date{\today}

\maketitle

\begin{abstract}

The late time behavior of waves propagating on a general curved
spacetime is studied.  The late time tail is not necessarily an
inverse power of time.  Our work extends, places in context, and
provides understanding for the known results for the Schwarzschild
spacetime.  Analytic and numerical results are in
excellent agreement.

\end{abstract}



\paragraph*{Introduction.}
Waves propagating on curved spacetime develop
``tails".  A pulse of gravitational waves (or other massless fields)
travels not only along the light cone, but also spreads out behind it,
and slowly dies off in ``tails" \cite{dewitt,price,gpp1,gpp2,cmp,leaver}.
This tail phenomenon is fascinating
theoretically, and has also been found in post-Newtonian
calculations to have possibly
observable secular effects on the phase of the
orbit of inspiralling binary systems\cite{poisson}.

For asymptotically late times this tail often
has a particularly simple behavior, namely, it decays as an inverse
power in $t$.  Detailed analyses
have been carried out for the Schwarzschild geometry, using both
analytic \cite {price,gpp1,cmp,leaver} and numerical techniques
\cite{price,gpp2}. These works are based on the Regge-Wheeler
perturbation formulation, in which the propagation of linearized
gravitational, electromagnetic and scalar waves is
described by the Klein-Gordon (KG) equation

\begin{equation}
D \phi(x,t) \equiv \left[ {\partial_{t}^{2}} - {\partial_{x}^{2}}
+V(x) \right] \phi (x,t) = 0  ,
\end{equation}

\noindent
with $V(x)$ describing the scattering of $\phi$
by the background geometry.
The late time tail has been explained in two different
ways: in terms of a branch cut in the Green's function in the
frequency plane \cite{leaver}, or in terms of scattering from large
radius in the Schwarzschild geometry \cite{price,gpp1}.

With the power-law tails in Schwarzschild spacetime well established
(although a thorough understanding, especially at time-like infinity,
is not complete\cite{private}), attention has been widened to more
general situations.  Tails in a Reissner-Nordstrom hole have been
studied\cite{gpp1}. Moreover, since
the late time tail comes from scattering at large radius
in the Schwarzschild case \cite{price}, it has been
suggested that power-law tails would develop even when there is no
horizon in the background \cite{gpp1}, implying that such tails should
be present in perturbations of stars, or after the collapse of a
massless field which does not result in black hole formation. In
\cite{gpp2}, the late time behavior of scalar waves evolving in its
own gravitational field or in gravitational fields generated by other
scalar field sources was studied numerically.  Power-law tails have
been reported for all these cases in numerical experiments, though
with exponents different from the Schwarzschild case.  These
interesting results call for a systematic study of the late time tail
in general nonvacuum and non-linear spacetimes.

There are several interesting questions in a general analysis.
Is the tail always a power in $t$?
What determines the asymptotic behavior --- the branch cut
in the Green's function
or the asymptotic form of the potential --- and how are the
two related?
What determines the magnitude and the time dependence of the tail, and
do these depend on local geometry and/or the presence of a horizon?

In this Letter, we study these questions using
(1) with a general $V(x)$.  Our work extends, and
places in context, the known results for specific background
geometries.

We shall first present numerical simulations of the late time behavior
for various $V(x)$, showing that the decay is not necessarily an inverse
power of $t$. Another interesting observation is that when the
parameters of the potentials are continuously varied, the behavior
of the late time tail can change discontinuously. We then
determine analytically the amplitude and the time-dependence of the
late time tail in terms of the strength of the branch cut in the
Green's function. The relation between the cut and the asymptotic
structure of the potential is obtained.
We establish that the local properties of the potential affect
only the magnitude but not the
time-dependence of the late time tail.

\paragraph*{Numerical Simulations.}
We study numerical evolutions of $\phi (x,t)$ given by (1) for
various $V(x)$.
The variable $x$ is related to, but not the same as,
the circumferential radius $r$ \cite{xr}.  For a nonsingular
metric, e.g., that of a star, $r \in (0, \infty)$ maps into $x \in (0,
\infty)$.  For a metric with an event horizon at $r = r_{o}$
(with $g_{tt} = 0$ at $r_{o}$ \cite{xr}), then  $r \in (r_{o},
\infty)$ typically maps into $x \in (-\infty, \infty)$ (the tortoise
coordinate).  The evolutions shown here are for the half line $x \in
(0, \infty)$ case, with boundary conditions $\phi =0$ at $x=0$ and
outgoing waves for $ x
\rightarrow +\infty$.  The full line case
[$x \in (-\infty, \infty)$, with outgoing wave boundary conditions for
$|x| \rightarrow \infty$] is basically the same.

In the following we consider two classes of
$V(x)$: potentials which go as a centrifugal barrier
$l(l+1)/x^2$ ( $l$ is an integer) plus $\overline{V}(x)$, with
$\overline{V}(x)$ being (i)
$x_o^{\alpha-2}/x^{\alpha} $ (``power-law
potentials'') or (ii) $(x_o^{\alpha-2}/x^{\alpha}) \log (x/x_o)$
(``logarithmic potentials'') when $x \rightarrow +
\infty$ for some $x_o$.
The logarithmic potential
includes the Schwarzschild metric as a special case.  The evolution is
basically independent (see below) of the initial data.  The cases
reported here use gaussian initial data for $\phi$ and $d \phi / dt$.

Fig 1. shows $ \log |\phi|$ versus $ \log t$ at some fixed point $x$
for several power-law potentials.
Solid lines represent the numerical evolutions; earlier times
are suppressed for clarity.  After some
quasinormal mode ringing, they
approach and coincide with the analytic results (to be
derived below)
representing power law decays $t^{- \mu}$,
where $\mu = 2l+\alpha$ except that in case (c)
$\mu$ jumps discontinuously to $2l+2 \alpha -2$.
Such jumps occur
whenever $\alpha$ is an odd integer less than $2l+3$.
(We assume throughout that the initial $d \phi / dt$ is not
exactly zero; otherwise the exponents $\mu$ increase by 1, also shown below.)

Logarithmic potentials
often lead to logarithmic late-time tails.  To
exhibit this behavior, Fig. 2 shows $ |\phi| t ^{2 l + \alpha }$ versus
$\log t$ for several logarithmic
potentials.  The numerical evolutions
approach and coincide with the analytic results
which represent decays in the form of $t^{-\mu} (\log t)^
{\beta}$, with $\mu = 2l+\alpha$, and $\beta = 1$,
except that $\beta$ jumps discontinuously to $0$ (and the
late-time tail becomes a power law) in case (c); such
jumps again occur when $\alpha$ go through an odd
integer less than $2 l + 3 $.

Table 1 summarizes these and other
cases studied but not shown here.  In all these examples,
$\alpha$ is taken to be larger than 2
($\alpha \le 2 $ will be discussed elsewhere).

\paragraph*{Analytic treatment.}
We first present a heuristic picture, and then state the necessary
modifications. Consider a wave
from a source point $y$ reaching a distant observer at $x$.  The
late time tail is caused by the wave first propagating to a point
$x' \gg x$, being scattered by $V(x')$, and then returning to $x$,
arriving at a time $t \simeq (x'-y)+(x'-x) \simeq 2x'$.
Thus at late times $ \phi \propto V(x') \simeq V(t/2)$.
In particular, if $V(x) \sim
x^{- \alpha} (\log x)^{\beta}$, then one expects the late time tail
to be $\sim t^{- \alpha} (\log t)^{\beta}$.

This picture requires two modifications.  First, a
centrifugal barrier,
corresponding to free propagation in 3 dimensions, does not
contribute to the late time tail, so that it is the remainder of the
potential, $\overline {V} (x)$,
that matters.  For $\overline{V}(x) \sim x^{-\alpha} (\log x)^{\beta}$,
late time tail turns out to be $t^{-(2l+ \alpha)} (\log t)^{\beta}$.
The suppression by a factor
$t^{-2l}$, at least in the case $\alpha = 3$, is known from studies of
specific black hole geometries \cite{price,gpp1}.
Secondly, if $\alpha$ is an odd integer
less than $2l+3$, the leading term in the late time tail
vanishes.  For $\beta=0$, the next leading
term is expected to be $t^{-(2l+2 \alpha -2)}$, while for $\beta=1$,
the next leading term is $t^{-(2l+\alpha)}$ without a $\log t$ factor.

Next we present a full analytic treatment for
the half-line problem; modifications for the
full-line case are straightforward.  The evolution of $\phi
(x,t)$ described by (1) is
\begin{equation}
\phi (x,t) = \int  dy \left[ G \dot{ \phi}(y,0)
+ \dot{G} \phi(y,0) \right]
\end{equation}
for $t>0$,
where the retarded Green's function $G(x,y;t)$ is defined by
$DG = \delta (t) \delta (x-y)$
with $G = 0$ for $t < 0$ and the outgoing
wave boundary condition as given for $\phi$ above.

The Fourier transform $\tilde{G}$ satisfies

\begin{equation}
\tilde{D}(\omega ) \tilde{G} \equiv \left[-\omega ^{2} -
\partial_{x}^{2} +V(x)\right] \tilde{G}(x,y;\omega ) = \delta (x- y)
\end{equation}

\noindent
and is analytic in the upper half $\omega$ plane. Define auxiliary
functions $f$ and $g$ by $\tilde{D}(\omega)
f(\omega,x)$ $=$ $\tilde{D}(\omega) g(\omega,x)$ $=$ $0$, where $g$ satisfies
$\lim_{x\rightarrow \infty} [e^{-i \omega x} g(\omega ,x)] = 1$,
and $f$ satisfies $f(\omega,x=0)$ = $0$, $f^\prime (\omega ,x=0)$ = $1$
\cite{fp} for the half-line problem, and $\lim_{x\rightarrow - \infty
} [e^{i \omega x} f(\omega ,x)] = 1$ for the full-line problem.  In
terms of $f$ and $g$, and henceforth assuming $y < x$,
$\tilde{G}(x,y;\omega )=
f(\omega ,y) g(\omega ,x) / W(\omega )$,
where the Wronskian $W(\omega ) = W(g,f) = g (df/dx)  - f (dg/dx)$
is independent of $x$.
Now express $G$ in terms of $\tilde{G}$ and for $t>0$, distort the
contour for the Fourier integral to a large semicircle in the
lower half $\omega$ plane.  One therefore identifies
three contributions to $G$, as follows.

First, the integral on the large semicircle
can be shown to vanish beyond a certain
time $t_{P}(x,y) = O(x)$, and does not affect the
late time behavior \cite{elsy}.

Secondly, at the zeros of the Wronskian $W(\omega )$ at $\omega  =
\omega _{j}, j = \pm 1, \pm 2, \ldots $ on the lower half
plane, $f$ and $g$ are proportional to each
other, and each of them satisfies {\it both} the left and the right
boundary conditions.  Hence they are quasinormal modes (QNM's), and
their collective contribution $G_{Q}$
vanishes exponentially at late
times.  [The case
where $G = G_{Q}$ (QNM's being complete) has been discussed in detail
\cite{elsy}.]

Finally, there may be singularities of $f$ and $g$
in $\omega$, which lead to the late time tail.  If the
potential $V(x)$ has finite support, say on $(0,a)$, then one can
impose the right boundary condition at $x = a^{+}$.  Integrating
through a {\it finite} distance with a non-singular
equation to obtain $g(\omega ,x)$
cannot lead to any singularity in $\omega$.  It is not
surprising that the same holds if $V(x)$ vanishes sufficiently rapidly
as $x \rightarrow +\infty$ \cite{elsy,poin}.  However, if $V(x)$ has
an inverse-power type tail, then
$g(\omega ,x)$ will have singularities
on the negative Im $\omega $ axis (see below),
in the form of a branch cut, as in the Schwarzschild case \cite{leaver}.
The cut extends to $\omega = 0$, and its tip controls the
late time behavior.  For the half-line problem, $f(\omega ,x)$ is
integrated from $x = 0$ through a finite distance, and hence does not
have any singularities in $\omega $.  For the full line case, $f$ is
dealt with in the same manner as $g$.  In all cases of interest,
the tail of $V(x)$ as $x \rightarrow -\infty$ is either faster than any
exponential, or precisely exponential.  For the former,
$f(\omega , x)$ has no singularities in $\omega$, while for the latter,
there will be a series of poles, but at a finite distance from
$\omega = 0$.  In either case, the spatial asymptotics as
$x \rightarrow -\infty$ has no bearing on the late time behavior.

It then remains to study
the spatial asymptotics as $x \rightarrow +\infty$, and the
consequent singularities of $g$.
First consider
a power law potential with $l = 0$. Applying the first Born approximation to
$\tilde{D}(\omega ) g = 0$ and starting with
$e^{i\omega x}$ as the zeroth order solution,
it is readily shown that
\begin{equation}
g(\omega ,x) \simeq e^{i\omega x} - I(\omega ,x)~,
\end{equation}
where
\begin{eqnarray}
I(\omega,x)
&&= \int_x^{\infty} dx' \ {\sin \omega( x-x') \over \omega} \
\overline{V}(x') e^{i \omega x'} \\
&&= {-e^{-i \omega x}
 \over (\alpha-1)} \left[ (-2i \omega x_o)^{\alpha-2}
\Gamma(2-\alpha) + \ldots \right] ~,
\end{eqnarray}
The omitted terms denote a convergent power series in $\omega$.
(This form is valid only for nonintegral $\alpha$; the integral case
can be obtained by taking a limit in the final result.)  The factor
$(-2i \omega x_o)^{\alpha -2}$ causes a cut in $\tilde{G}$
on the negative Im $\omega$ axis. Thus, we have
\begin{equation}
G(x,y;t \rightarrow \infty)
\simeq - {f(0,x) f(0,y) \over g(0,0)^2}
{2^{\alpha-1} x_{o}^{\alpha-2} \over t^{\alpha}}
\end{equation}
The late time behavior
is therefore $\phi \simeq t^{- \alpha}$ generically
[unless $\dot{\phi}(y,t=0) =0$].

Notice that the time dependence is determined solely by the asymptotic
form of $V$, while the magnitude, involving $f$ and $g(0,0)$, is
sensitive to
the local geometry, and hence to the existence or otherwise
of a horizon (in the full line case).  Note also that the Born
approximation, which is strictly valid at very large $x$, has been
used only to evaluate an $x$-independent Wronskian; therefore
the results are exact at large $t$.

Next consider $ l \neq 0$.
It is necessary to handle the centrifugal barrier exactly, and
to treat only $\overline{V}(x)$ using the Born approximation.  The zero
order solutions are now Hankel functions rather than
plane waves.  A somewhat lengthy calculation, along the same
lines as before, then leads to
\begin{equation}
G(x,y;t \rightarrow \infty) \simeq
- {f(0,x) f(0,y) \over g_o^2}
{C(l,\alpha) F(\alpha) \over t^{2l+\alpha}}
\end{equation}
where
\begin{equation}
C(l,\alpha) =
\prod_{j=0}^{l-1} { \alpha-2j-3 \over \alpha+1+2j} \ ,
\qquad l = 1, 2, \ldots
\end{equation}
and $C(l, \alpha) = 1$ for $l = 0$, with $F( \alpha ) = 2 (2
x_o)^{\alpha-2}
\Gamma(2l+\alpha)/ \Gamma( \alpha )$, and
$g_o \equiv \lim_{\omega \rightarrow 0} [(i \omega)^l W(g,f)]$,
which is finite \cite{poin} and reduces to $g(0,0)$ for $l = 0$.
The extra power of $\omega ^l$ in the definition of $g_o$ is
responsible for the suppression of the late time tail by an
extra factor of $t^{-2l}$, so that
in general [unless $\dot{\phi}(y,t=0)$ vanishes],
$\phi(x,t) \sim t^{-(2l+\alpha)}$ at late times.
But there is an exceptional case: when
$\alpha$ is an odd integer less than $2l+3$, $C(l,\alpha ) = 0$ and
the late time tail vanishes in first order
Born approximation; higher order approximations
representing multiple scatterings from asymptotically far
regions give the
next term going as $\sim t^{-(2l + 2 \alpha - 2)}$.

Generically the late time behavior is linear in the
potential (first Born approximation).
By applying $(-\partial/
\partial \alpha)$ on the corresponding power-law potential, we obtain
from (8) that for logarithmic potentials:
\begin{equation}
G(x,y;t \rightarrow \infty) \simeq
{f(0,x)f(0,y) \over g_o^2}
{\partial \over \partial \alpha} \left[{C(l, \alpha)
F(\alpha) \over t^{2l+\alpha}} \right]
\end{equation}
The leading terms are $t^{-(2l + \alpha)} (c \log t +d)$,
except that
$c \propto C(l,\alpha)$ vanishes when $\alpha$
is an odd integer less than $2l+3$.
The Schwarzschild case ($\alpha = 3$) with $l
\ne 0$ belongs to this exception;
as is well known \cite{price,gpp1,gpp2,cmp,leaver},
the late time behavior is a power-law with exponent
$-(2l+3)$, and no $\log t$ factor.

Analytic results for $\phi$ at large $t$
can then be obtained from (7) and (8) with {\it no}
adjustable parameters (except for case (c) in Fig. 1; see below),
and are plotted as dashed lines in Figs. 1 and 2.
The agreement is perfect.
The exceptional cases (c) deserve mention.  For case (c) in
Fig. 1, the leading term vanishes, and the dashed line shown
represents the next leading term arising from multiple
scattering, whose time dependence is determined,
but whose magnitude has been left
as an adjustable normalization.  For case (c) in Fig. 2,
the vanishing of the leading term implies that the asymptotic
slope should be zero (i.e. no log $t$, but only a pure power,
whose magnitude is determined),
and this indeed agrees with the numerical results,
with no adjustable parameters.
These results are for
time-like infinite.  Results for null-infinity will
be given in detail elsewhere.

In conclusion we have achieved an analytic understanding
of the late time tail in such systems, with asymptotic
formulas agreeing perfectly with numerical results.  The
late time behavior is due to the tip of the cut in the
frequency plane, which arises from scattering at large
radius.  For a potential that is a centrifugal barrier
plus $\sim
x^{- \alpha} (\log x)^{\beta} (\alpha > 2, \beta = 0,1)$,
the late time tail is generically
$\sim t^{- (2l+\alpha)} (\log t)^{\beta}$.  The possibility
of a logarithmic factor in the leading late time behavior appears
not to be widely known.  Moreover, the case where $\alpha$
is an odd integer less than $2l+3$ is exceptional, and interestingly
enough the most familiar Schwarzschild case belongs to
this exception.

We thank Richard Price for discussions.
We acknowledge support from the Croucher Foundation. WMS is also
supported by the CN Yang Visiting Fellowship and
the US NSF (Grant No. 94-04788).

\newpage
\centerline{\bf FIGURE CAPTIONS}

\noindent Fig. 1 \ $\log |\phi|$ versus $\log t$,
 for several power-law potentials.  (a) $l = 0, \alpha = 3$, (b) $l = 1,
\alpha = 2.9$; (c) $l = 1 , \alpha = 3 $; (d) $l = 1 , \alpha = 3.1 $.
Solid lines are numerical evolutions from generic initial data while
the dashed lines are analytical results.  They are indistiguishable
for $\log t > 8$.  To make the four sets of lines stagger,
vertical shifts have been applied: (b) downwards by 6.0,
(c) downwards by 12.0, (d) downwards by 23.0.

\noindent Fig. 2 \
$ K | \phi | t^{2l + \alpha}$ versus $\log t$ for several logarithmic
potentials (a) $l = 0$, $\alpha = 3$, (b) $l = 1$, $\alpha = 2.9$;
(c) $l = 1$, $\alpha = 3 $; (d) $l = 1$, $\alpha = 3.1 $.
Solid lines are numerial results while dashed lines are
analytic results. They are indistiguishable for $\log t > 9.5$.
For clarity, the data are multiplied by a constant $K$
with (a) $K = 10^{-9}$ (b) and (c) $K = 5.6 \times 10^{-10}$
and (d) $K= 5.86 \times 10^{-10}$.


\newpage

\vskip 4in
\renewcommand {\arraystretch}{1.76}

\begin{tabular}{|c|c|c|c|} \hline
$\overline{V}(x), x \rightarrow \infty$ &
$\alpha > 2$ & $ \phi (t), t \rightarrow \infty $ \\ \hline
$x_o^{\alpha -2}/x^{\alpha} $ & odd integer $< 2l+3$ &
$t^{- \mu}, \mu > 2l + \alpha$  \\ \cline{2-3}
& all other real $\alpha$ &  $t^{- ( 2l + \alpha )} $ \\ \hline
$x_o^{\alpha-2}/ x^\alpha \log (x / x_o) $
& odd integer $< 2l+3$ & $t^{- (2l + \alpha)} $ \\ \cline{2-3}
& all other real $\alpha$ & $ t^{- ( 2l + \alpha ) } \log t$
\\ \hline
\end{tabular}
\vskip 25pt
\noindent Table 1 \ Behavior of late time tails for potentials
going as $l(l+1)/x^2 + \overline{V}(x)$ when $x \rightarrow \infty$.
\hoffset=0.in


\begin{references}
\bibitem{xr} For example, in a static spherically symmetric
background, the evolution of linearized waves is given exactly by the
KG equation (1) with the variable $x$ related to the circumferential
radius $r$ by $-g_{tt}(dx)^{2} = g_{rr}(dr)^{2}$, with $t$ the Killing
time.
\bibitem{fp} We assume here $f'$ is nonzero at the origin.
Otherwise, if $f \sim x^{l+1}$ as $x \rightarrow 0$, then we take
$\lim_{x \rightarrow 0} x^{-(l+1)} f(\omega,x) = 1$ instead of $f'=1$.
\end{references}
\end{document}